\documentclass[prb,twocolumn,showpacs]{revtex4}
\usepackage{amsmath}
\usepackage{amssymb}
\usepackage{bm}
\usepackage{epsfig}
\usepackage{graphicx}
\usepackage{natbib}
\usepackage{color}
\usepackage[T2A]{fontenc}
\usepackage{enumitem}

\newcount\timehh  \newcount\timemm
\timehh=\time \divide\timehh by 60
\timemm=\time
\begin{document}

\title{Magnetic field induced valence band mixing \\in [111] grown
 semiconductor quantum dots}

\author{M.~V.~Durnev}
\author{M.~M.~Glazov}
\author{E.~L.~Ivchenko}

\affiliation{Ioffe Physical-Technical Institute of the RAS, 194021  St.-Petersburg, Russia}

\author{M.~Jo}
\author{T.~Mano}
\author{T.~Kuroda}
\author{K.~Sakoda}

\affiliation{National Institute for Material Science, Namiki 1-1, Tsukuba 305-0044, Japan}

\author{S.~Kunz}
\author{G.~Sallen}
\author{L.~Bouet}
\author{X.~Marie}
\author{D.~Lagarde}
\author{T.~Amand}
\author{B.~Urbaszek}

\affiliation{Universit\'e de Toulouse, INSA-CNRS-UPS, LPCNO, 135 Avenue Rangueil, 31077 Toulouse, France}

\date{\today, file = \jobname.tex, printing time =
\number\timehh\,:\,\ifnum\timemm<10 0\fi \number\timemm}

\begin{abstract}
We present a microscopic theory of the magnetic field induced mixing of heavy-hole states $\pm 3/2$ in GaAs droplet dots grown on (111)A substrates. The proposed theoretical model takes into account the striking dot shape with trigonal symmetry revealed in atomic force microscopy. Our calculations of the hole states are carried out within the Luttinger Hamiltonian formalism, supplemented with allowance for the triangularity of the confining potential. They are in quantitative agreement with the experimentally observed polarization selection rules, emission line intensities and energy splittings in both longitudinal and transverse magnetic fields for neutral and charged excitons in all measured single dots.
\end{abstract}

\pacs{72.25.Fe, 73.21.La, 78.55.Cr, 78.67.Hc}

\maketitle

\section{Introduction}
Thorough understanding and tuning of the optical properties of semiconductor quantum dots (QDs) has allowed the development of efficient quantum emitters.\cite{Tartakovskii:2012a} In parallel, the electrical and optical manipulation of the spin states of single carriers in QDs has made unprecedented control over quantum states in the condensed matter possible.\cite{Petta:2010a,Latta:2011b} Single spin manipulations and quantum optics with single photons are directly related to the optical selection rules.\cite{Dyakonov:2008a} The band structure given by the symmetry and strength of the carrier confinement potentials defined, e.g., by the dot shape and its chemical composition plays a crucial role in defining the exact carrier spin state and, correspondingly, the polarization of the emitted and absorbed photons. 

\begin{figure}[hptb]
\includegraphics[width=0.49\textwidth]{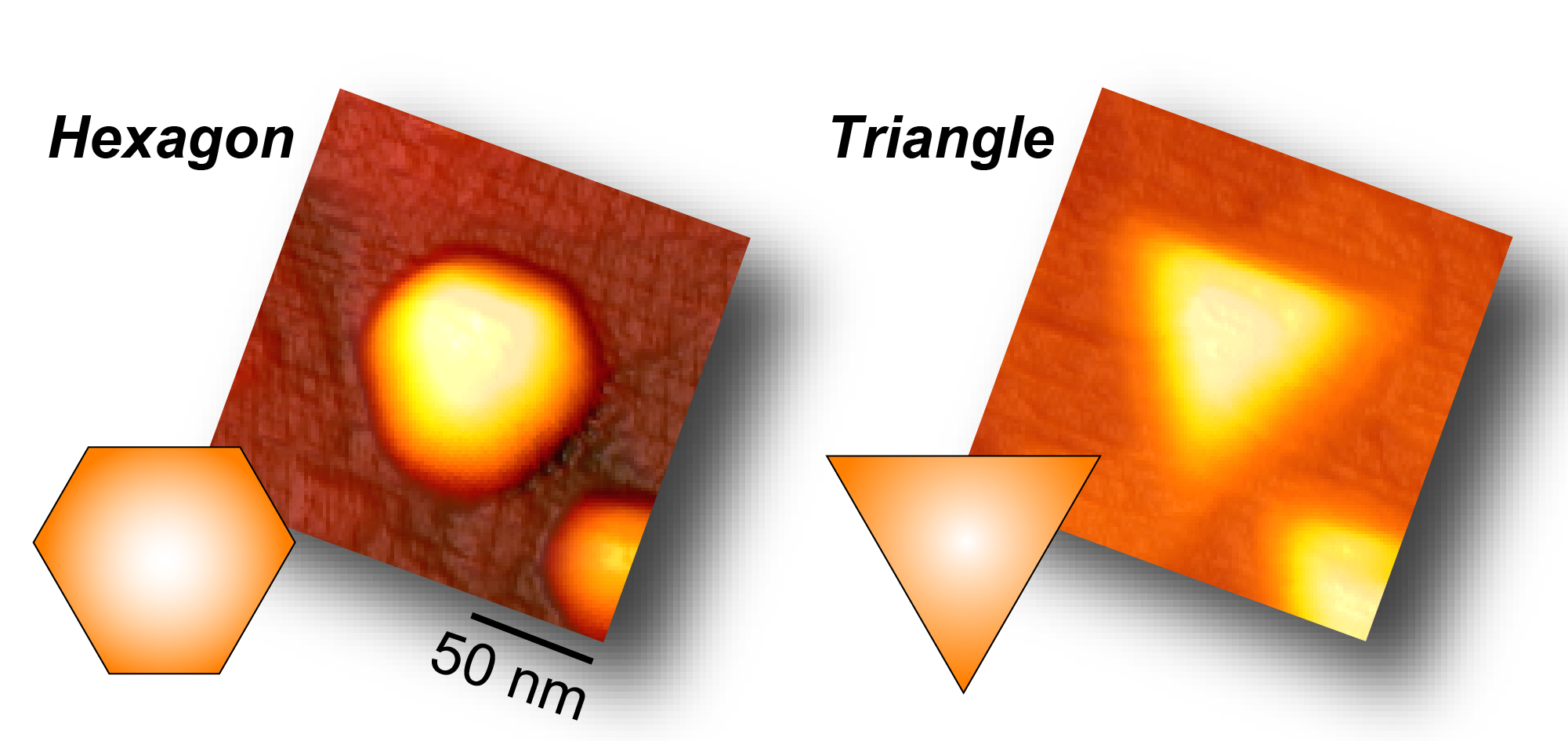}
\caption{\label{fig:fig1} (Color online):  AFM images of GaAs quantum dots grown on (111)A substrates reveal the triagonal symmetry of the dots, with shapes varying from irregular hexagons to equilateral triangles depending on crystallisation temperature.
}
\end{figure} 

The majority of devices based on self-assembled quantum dots is grown along the standard [001] crystal axis, for which the dots tend to have an elliptic or rectangular shaped base. The associated C$_{2v}$ point symmetry results in a complex exciton fine structure \cite{Bayer:2002a,Poem:2010a,ivchenko97pss} characterized by well-defined linearly polarized eigenstates being split in the case of neutral X$^0$ exciton emission by the electron-hole exchange interaction. The main aim here is the suppression of this fine structure splitting to allow the emission of entangled photon pairs, as convincingly demonstrated  using a variety of techniques.\cite{Stevenson2006,Plumhof2012,Langbein2004,Akopian2006} 
Such a low, C$_{2v}$ point symmetry is also relevant for the [001]-grown quantum dots with circular or square-shaped base provided that the top and bottom interfaces are non-equivalent or if an in-plane strain is present.\cite{ivchenko97pss,bester05, koudinov04,zunger2011} A promising, alternative approach to suppress the fine structure splitting of the neutral exciton is the growth along the [111] crystal axis, which is also the orientation of most nano-wires.\cite{Lu2006} This growth axis has the advantage of providing in principle quantum-dot shape of the higher C$_{3v}$ point symmetry. Really,
small fine structure splittings in \textit{as grown} [111] quantum dot
structures have been recently predicted \cite{Schliwa2009,Singh2009} and observed \cite{Mano2010,Karlsson2010,Stock2010,Treu:2012a} making such structures a very promising system for entangled photon pair emission.

\begin{figure}[h!]
\includegraphics[width=0.45\textwidth]{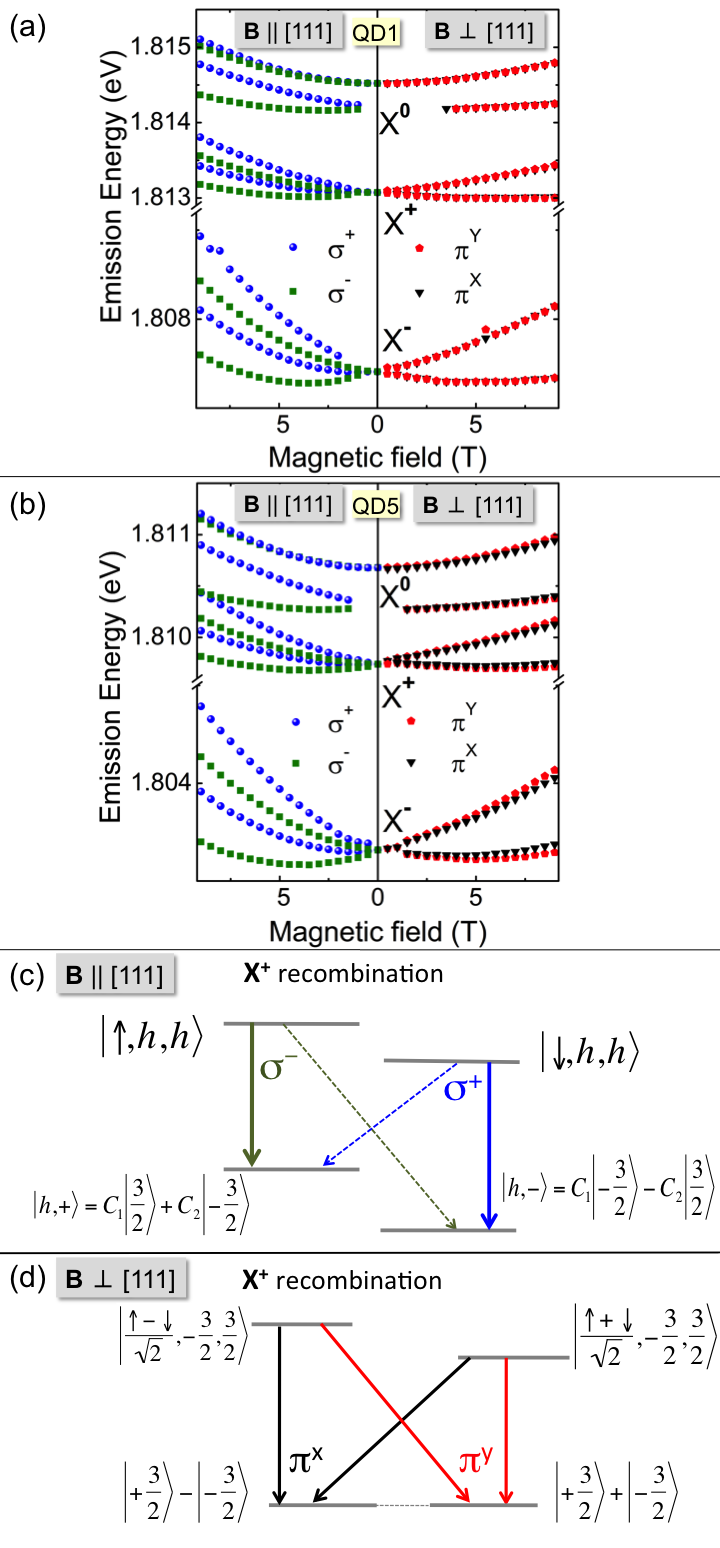}
\caption{\label{fig:fig2} (Color online) (a) QD1 Left: Measured transition energies as a function of the applied magnetic field $\bm{B} \parallel [111]$. Right: Measurements for $\bm{B} \perp [111]$. (b) Same as (a), but for QD5. (c) Polarization selection rules for X$^+$~recombination in the Faraday geometry ($B \parallel [111]$) with
  mixed hole states, see Eq.~(\ref{2x2}), where $C_{1,2}$ are coefficients determining hole eigenstates.  
  (d) Polarization selection rules for X$^+$~recombination in the Voigt geometry ($B \perp [111]$).
}
\end{figure}

The exact electron spin states, selection rules and optical properties in any semiconductor nanostructure are primarily determined by the symmetry of the investigated system.\cite{Ivchenko1995} As compared to the well studied [001] grown quantum dots, the triagonal symmetry associated to the growth along the [111] axis revealed in the Atomic Force Microscopy (AFM) images presented in Fig. \ref{fig:fig1}, gives rise to surprising new spin effects discovered recently and studied in detail in the present paper: \\
\mbox{}\hspace{5 mm} (i) In a magnetic field applied along the [111] growth axis (Faraday geometry), the neutral and singly charged exciton complexes show a total of four interband transitions each, two $\sigma^+$ and two $\sigma^-$ polarized,\cite{PhysRevLett.107.166604}~see Fig.~\ref{fig:fig2}a and b, left panels.\\
\mbox{}\hspace{5 mm} (ii) In a transverse magnetic field applied perpendicularly to the growth axis (Voigt geometry), the charged exciton emission is split into two lines, each doubly degenerate, see Fig.~\ref{fig:fig2}a and b, right panels. Experiment shows that the heavy hole states have zero Zeeman splitting in the transverse field.

Both of these findings, (i) and (ii), detailed in section \ref{sec:exp} are in stark contrast to results obtained in samples grown along the [001] axis (see, for example, Refs.~\onlinecite{Bayer:2002a,Leger:2007a,Xu:2007a,Krebs:2010a}), where two transition lines are observed in the Faraday, and four well separated transitions in the Voigt geometry. Symmetry arguments, presented in Sec.~\ref{sec:phenomen}, allow us to identify the key ingredients needed to explain the characteristic interband transitions: the valence heavy hole pseudo spins $|+3/2 \rangle$ and $|-3/2 \rangle$ are coupled in longitudinal magnetic field, whereas they remain uncoupled in transverse field. Hence, the magnetic field induced tuning of the energy level structure and the underlying symmetry in [111] grown quantum dots, see Fig.~\ref{fig:fig2}c and d, makes it possible to develop new, original approaches to coherent electron and hole spin control.\cite{press08,Benny:2011a,Godden:2012a}

The main part of this paper, Sec.~\ref{sec:micro}, is devoted to a detailed microscopic model that provides quantitative evaluation of the magnetic field induced heavy-hole mixing in the Faraday geometry. 
We show that the main contribution to the heavy-hole mixing results from the influence of the triangular quantum dot shape on the envelope part of the valence states wave-functions, rather than the underlying lattice symmetry due to the [111] growth axis.
Details of the heavy hole Zeeman effect calculations are provided in the Appendix. 

The experimental and theoretical results presented so far open up for the true quantum mechanical engineering of the Zeeman effects in nanostructures.

\section{Experimental Results}
\label{sec:exp}

The samples were grown by droplet epitaxy using a conventional molecular
beam epitaxy system \cite{Mano2010,Abbarchi2010,Belhadj2008} on a GaAs(111)A substrate. 
The GaAs dots are grown on 100~nm thick Al$_{0.3}$Ga$_{0.7}$As  barriers and are covered by 50~nm of the same material.
The barrier and the quantum dot material have the same lattice parameter, \cite{Keizer:2010a} the samples are within good approximation strain free.
As a result, there are no piezo-electric effects to take into account, despite the [111] growth direction.
There is no continuous wetting layer in the sample connecting the
dots, whose typical height is $\simeq$~3~nm, and radius is $\simeq$~15~nm, see details in Ref.~\onlinecite{Mano2010}, which results in strong confinement of the carriers in the dots.
Atomic force microscopy studies carried out on similar samples grown on  a GaAs(111)A substrate reveal a dot base with trigonal symmetry, see Fig.~\ref{fig:fig1}, that evolves from irregular hexagons to equilateral triangles as the crystallization temperature is varied.\cite{Jo:2012a}
Single dot photoluminescence (PL) at 4~K is recorded with
a home build confocal microscope with a detection spot diameter of
$\simeq 1$~$\mu$m. The detected PL polarization is analysed, the
signal is dispersed by a spectrometer and detected by a Si-CCD
camera. Optical excitation is achieved by pumping the AlGaAs barrier
with a HeNe laser at 1.96~eV that is linearly polarized to exclude the
effects of optical carrier orientation and dynamic nuclear
polarization.\cite{belhadj09, urb_arxiv}

Three exciton complexes dominate the PL emission at zero applied field: The already mentioned neutral X$^0$, the negatively charged exciton (trion) X$^-$ comprised of 2 electrons in a spin singlet state and 1 hole, and the positively charged trion X$^+$ (1 electron, 2 holes in a spin singlet state) are identified using fine structure analysis and
optical orientation experiments.\cite{belhadj09} The high symmetry of the dots grown along [111] is
reflected in typical values for the splitting of the X$^0$ emission
due to anisotropic electron-hole exchange $\delta_1$ of a few $\mu$eV, \cite{Mano2010} extracted from angle dependent PL polarization
analysis in the linear basis.  

The X$^+$ and X$^-$ trions of all dots investigated exhibit four transitions in longitudinal magnetic fields, as seen from the emission spectra of two typical dots presented in Fig.~\ref{fig:fig2}. Surprisingly, there are clear polarization selection rules: two transitions are $\sigma^-$ polarized, the two others $\sigma^+$. 
The competition between the exchange interaction and magnetic field effects leads to intriguing observations for the neutral exciton X$^0$: The Zeeman splitting is strongly non-monotonous,\cite{PhysRevLett.107.166604} see Fig.~\ref{fig:QD6} for a striking example. 

Figure \ref{fig:fig2} shows also the spectra of the same two typical QDs in the Voigt geometry: For QD1 shown in panel (a) the total number of lines we are able to resolve is two for each charged exciton (X$^+$ and X$^-$), and not four as could be expected by analogy with [001] samples.\cite{Bayer:2002a,Leger:2007a,Xu:2007a,Krebs:2010a} This is true both in the circular ($\sigma^+/\sigma^-$) and the linear basis ($\pi^x/\pi^y$). Analysing a quantum dot with lower symmetry allowed us to uncover the true polarization basis and lift the remaining degeneracies. Indeed, the shape of QD5 probably deviates from an ideal pyramid with an equilateral triangular base and each line of the initially observed doublets undergoes a further very small splitting with the increasing field. According to Fig.~\ref{fig:fig2}b at 9~T we observe four resolution limited transitions in the linear basis for the X$^+$ and X$^-$ trions with the additional splitting $\Delta E_{\rm max} \approx 40~\mu$eV. As shown below the presence of four transitions in the Voigt geometry indeed corresponds to the non-zero in-plane heavy hole $g$-factor allowed for (111) structures with point symmetry lower than $C_{3v}$. 

Moreover, both in the Faraday and the Voigt geometries nominally dark exciton (with total spin projection $m_z=s_z + j_z = \pm2$, where $s_z$ and $j_z$ are the electron and hole spin components) gains an oscillator strength, allowing a precise determination of the dark-bright X$^0$ splitting in the range of 300 to 400~$\mu$eV (determined for several tens of QDs), which is strongly enhanced, due to confinement, compared to the GaAs bulk value and the larger GaAs interface fluctuations dots.\cite{Gammon:2001a}

\section{Theory}
\subsection{Symmetry Analysis} \label{sec:phenomen}

From the analysis of the single dot PL in linear basis, see Ref. \onlinecite{belhadj09} for details, we are able to extract an X$^0$ fine structure (anisotropic) splitting $\delta_1$ of typically only a few $\mu$eV. This, together with the analysis of possible symmetries for zinc-blende lattice based structures grown along $z\parallel[111]$ axis, allows us to conclude that the investigated QDs  have (close to ideal) trigonal $C_{3v}$ point symmetry, in perfect agreement with the AFM measurements shown in Fig.~\ref{fig:fig1}. 

In what follows we introduce the Cartesian coordinate frame $x \parallel
[11\bar{2}]$, $y \parallel [\bar{1}10]$ and $z \parallel [111]$  relevant for the system under study and analyze the effect of a longitudinal magnetic field along the growth axis $z$. For a nanostructure of the C$_{3v}$ symmetry we assume one of the planes $\sigma_v \in {\rm C}_{3v}$ to contain the axes $x$ and $z$. The phases of heavy-hole basis functions  $| 3/2 \rangle, |-3/2\rangle$
are chosen to show the same point-group symmetry properties as the functions $-$$\uparrow (x + {\rm i} y)$ and $\downarrow (x - {\rm i} y)$, respectively, where $\uparrow, \downarrow $ are the spin-up and spin-down columns. 

For dots with the C$_{3v}$ symmetry, the group theory predicts a magnetic field
induced mixing between the heavy hole states $| 3/2 \rangle$ and $|-3/2\rangle$.
These two states form a reducible representation reduced to the two one-dimensional representations $\Gamma_5$ and $\Gamma_6$
with the basis states $| \Gamma_5 \rangle = | 3/2 \rangle + {\rm i} | - 3/2 \rangle$ and $
| \Gamma_6 \rangle = | 3/2 \rangle - {\rm i} | - 3/2 \rangle$. The latter functions are related by the time inversion operation and, therefore, at zero magnetic field have the same energy. Since the direct product
\begin{equation} \label{dirprod}
(\Gamma_5 + \Gamma_6) \times( \Gamma^*_5 + \Gamma^*_6) = 2\Gamma_1 + 2\Gamma_2
\end{equation} 
contains two representations $\Gamma_2$ according to which transforms the magnetic field component $B_z$, 
the heavy-hole Zeeman splitting in the basis $| 3/2 \rangle, |-3/2\rangle$ is described by
a 2$\times$2 matrix with two linearly independent parameters 
\begin{equation} \label{2x2}
{\cal H}_{\bm B} = \frac12\ \mu_B B_{z} \left[ \begin{array}{cc}
g_{h1} & g_{h2}\\  g_{h2} & - g_{h1}\end{array} \right] \:. 
\end{equation}
Here $\mu_B$ is the Bohr magneton, $g_{h1}$ and $g_{h2}$ are the
effective hole $g$-factors, the latter also being real if the trigonal symmetry includes the mirror reflection in the plane $(x,z) \parallel (\bar{1}10)$. The similar result holds for the trigonal point groups D$_3$ and {D$_{3d}$ = D$_3 \times {\mathrm C}_i$}. 
For the low-symmetry trigonal C$_3$ and {S$_6$ = C$_{3} \times{\mathrm C}_i$} point groups, the mixing of the heavy holes by the longitudinal field is allowed as well. In this case, however, the symmetry does not fix the orientation of lateral axes $x$ and $y$ where the off-diagonal matrix elements are real. 
For all hexagonal groups including C$_{3h}$ and D$_{3h}$ the longitudinal field-induced mixing of the heavy hole states is forbidden, and the value of $g_{h2}$ in Eq.~\eqref{2x2} vanishes. This is also true for the orthorhombic C$_{2v}$ group and tetragonal D$_{2d}$ group which reproduce the symmetry of the conventional [001] grown dots (lens- and disk-shaped, respectively).

For completeness we note that the experiments in the Voigt geometry shown in the right panel of Fig. \ref{fig:fig2}a and b infer that heavy holes are not mixed by the in-plane magnetic field. Indeed, the direct product (\ref{dirprod}) does not contain the representation $\Gamma_3$ describing the symmetry properties of the field components $B_x$ and $B_y$ and, therefore, the C$_{3v}$ point group does not allow for the mixing of the $\pm 3/2$ states by the in-plane magnetic field. This result is as well valid for
all 12 point groups of the hexagonal crystal family including the hexagonal and trigonal crystal systems. On the other hand, the mixing is allowed by the C$_{2v}$ or D$_{2d}$ point-group symmetry where small but reliably measured in-plane heavy-hole $g$-factors were observed.\cite{Mar99}

\subsubsection{Charged excitons X$^\pm$} 
In a longitudinal magnetic field, the hole eigenenergies are
$E_{\pm} = \pm g_h \mu_B B_{z}/2$ with $ g_h = \sqrt{g_{h1}^2 + 
g_{h2}^2}$ and the hole eigenstates $|h,\pm \rangle$ are admixtures
of $|3/2\rangle$ and $|-3/2\rangle$, as indicated in Fig.~\ref{fig:fig2}c, with
the coefficients  
$C_{1,2}$ determined solely by the ratio $g_{h2}/g_{h1}$, see Ref.~\onlinecite{PhysRevLett.107.166604} for details.  
For non-zero $g_{h2}$, all the four radiative transitions are allowed,
each transition being  
circularly polarized, either $\sigma^+$ or $\sigma^-$.
The transition energies include a Zeeman contribution as well as a diamagnetic shift common for all lines. The splitting between different spin states is determined by combinations of the
electron and hole effective $g$-factors which allows to find a pair of
parameters, $g_e$ and $g_h=\sqrt{ g_{h1}^2 + g_{h2}^2}$. The
intensities of circularly-polarized lines are proportional to
$C_1^2$ and $C_2^2$ and independent of the magnetic field, in full agreement with our
experiments. From the ratio of intensities of identically-polarized lines
we can find the ratio $g_{h1}/g_h$ and, therefore, determine values of
$g_e$, $g_{h1}$ and the absolute value of $g_{h2}$.~\cite{PhysRevLett.107.166604}

\begin{table}
\caption{\label{tab:table1} $g$-factors (typ. error $\leq 10 \%$) for charged
  and neutral excitons obtained from fitting the data. For the X$^0$ the
  $g_e$ and $g_h$ values obtained for X$^+$ for the same dot are taken and only
 $|g_{h2}|$ is varied to fit the bright and dark $X^0$ splitting
  \textit{simultaneously}. } 

\begin{ruledtabular}
\begin{tabular}{ccccc}  
& QD1 & & QD5 \\
  geometry & Faraday & Voigt & Faraday & Voigt   \\
\hline
X$^-$ : $g_{e}$ & 0.48 &	0.80 &	0.49 & 0.73  \\
$g_{h}$ & 0.79 &	0	& 0.83 & 0.07  \\
$|g_{h2}|$ & 0.57 &	- &	0.53 & - \\
\hline
X$^+$ : $g_{e}$  & 0.47 & 0.84 & 0.47 &	0.79	 \\
$g_{h}$ & 0.72 & 0.03 & 0.71 & 0.07 \\
$|g_{h2}|$ & 0.7	& - & 0.62 & -  \\
\hline
X$^0$ : $|g_{e}|$ & 0.47 & 0.80 & 0.47 & 0.77  \\
$g_{h}$ & 0.72 & 0 & 0.71 & 0.07 \\
$|g_{h2}|$ & 0.59	& - & 0.5 & -  \\
\end{tabular}
\end{ruledtabular}
\end{table}

\subsubsection{Neutral exciton X$^0$}

When analysing the emission energies and polarization of the neutral exciton X$^0$, a competition between the magnetic field and the electron-hole exchange interaction has to be considered. Taking into account that in the absence of magnetic field the bright exciton states with the angular momentum components $m_z = \pm 1$ are separated from the dark states with $|m_z|=2$ by the energy $\delta_0$, we obtain for the X$^0$ sublevel energies\cite{aniso111, PhysRevLett.107.166604}
\begin{eqnarray} \label{esm}
E_{s_{{z}},{\tilde j}} = s_{{z}} g_e \mu_B B_{z} + \frac12 (\delta_0 + {\tilde j}\delta_{s_{{z}}})\:,  \hspace{1.5 cm} \\
\mbox{} \hspace{0.5 cm} \delta_{s_{{z}}} = \sqrt{\delta_0^2 + (g_h\mu_B B_{z})^2 - 4 s_{{z}} g_{h1} \mu_B B_{z} \delta_0}\:.\nonumber
\end{eqnarray}   
Here $\tilde j = \pm 1$ labels heavy-hole eigenstates in the presence of magnetic field and $\delta_0>0$. It follows from Eq.~\eqref{esm} that the splitting of bright X$^0$ states, $E_{+1/2,+{1}}(B_{z}) - E_{-1/2,+{1}}(B_{z})$,  can be a
non-monotonous and even sign-changing function of $B_{z}$ as has been confirmed experimentally.\cite{PhysRevLett.107.166604} This
result is another striking difference when comparing with the [001] grown dots, where the
observed splitting increases monotonously as a function of the applied
longitudinal field.\cite{Bayer:2002a, Leger:2007a}

\begin{figure}[hptb]
\includegraphics[width=0.47\textwidth]{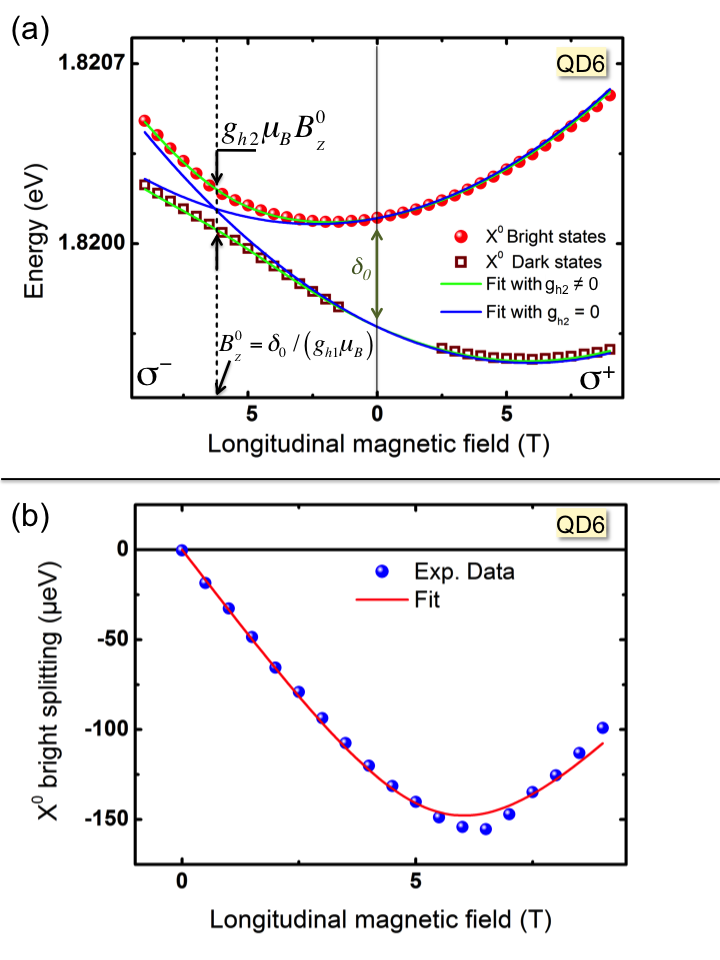}
\caption{\label{fig:QD6} (Color online): QD6 (a) Transition energy of the nominally dark (squares) and bright (circles) neutral exciton X$^0$ state as a function of the applied magnetic field (Faraday geometry). Left- and right-hand sides represent the states with $s_z = 1/2$ and $s_z = - 1/2$, respectively. Fit (green curve) of the data using Eq. \ref{esm} allows to extract $\delta_0=422$~$\mu$eV, $g_h=1.16$, $g_e=0.49$, $g_{h1}=1.05$ and $g_{h2}=0.42$. The fit included the diamagnetic shift for both curves as $E = E_0 + \alpha B_z^2$ with $E_0= 1.819676$~eV and $\alpha = 4.15$~$\mu$eV/T$^2$. Blue curve:  identical parameters apart from $g_{h2}=0$. (b) X$^0$ exciton bright splitting as a function of the magnetic field. Solid line is the fit obtained by using Eq. \ref{esm}.
}
\end{figure} 

As a particular example we present in Fig.~\ref{fig:QD6} the transition energies of QD6 determined in the Faraday geometry. This dot shows comparatively weak heavy hole mixing, resulting in a relatively small value of  $g_{h2}=0.42$ compared to $g_h=1.16$, and allows us to illustrate how well Eq.~(\ref{esm}) fits the experiments: We even observe an anti-crossing of dark and bright X$^0$ states, see vertical dashed line in Fig. \ref{fig:QD6}a. Here the consequences of introduction of $g_{h2}$ into the Zeeman Hamiltoninan, Eq.~\eqref{2x2}, and, hence, into the fit are striking. First, if $g_{h2}$ is set artificially to zero, there is no anti-crossing but a clear crossing. This crossing appears at the characteristic field $B_z^0=\delta_0/(g_{h1}\mu_B)$. The dark-bright separation at this field $B_z^0$ is exactly $|g_{h2}\mu_B B_z^0|$. Figure~\ref{fig:QD6} shows directly the evolution of the dark and bright X$^0$ transitions in the presence of heavy hole coupling, i.e., of non-zero $g_{h2}$. Second, the small value of $g_{h2}$ leads to a strongly non-monotonous bright exciton splitting going from 0 down to $-$150~$\mu$eV at 6.5 T back up to $-$100~$\mu$eV at 9 T. The applied magnetic fields are too low here to see a reversal of the sign, but the tendency is clear\cite{nonmonZ,Jovanov:2012a} (we observed, however, this sign reversal in other dots, see Ref.~\onlinecite{PhysRevLett.107.166604}). Hence, the high quality of the fits of the neutral exciton splitting in applied fields as presented in Fig.~\ref{fig:QD6} underlines the important role of heavy hole coupling, i.e., non-zero $g_{h2}$. 

It is worth stressing that all sets of experimental data obtained in the Faraday geometry for several tens of quantum dots, both for trion and exciton emission, is well reproduced within a simple model, where linear-in-$B_z$ contributions to the Zeeman effect with $g_{h1}$ and $g_{h2} \ne 0$ are taken into account only.\cite{diamagnB2,PhysRevB.85.155305,PhysRevLett.107.127403}
Thus, an introduction of only one additional parameter $g_{h2}$ allows us to fit all the transition energy splittings of the neutral and charged excitons, i.e., explain all the differences between the excitonic transitions observed in the [111] grown droplet dots and standard dots grown along [001]. In the symmetric [111] grown dots under study we often find strong heavy hole mixing as $|g_{h2}|>|g_{h1}|$. Such surprisingly high values of the off-diagonal $g$-factor are consistent with the microscopic theory developed in the next Subsection. 

\subsection{Microscopical theory}
\label{sec:micro}
In Ref.~\onlinecite{PhysRevLett.107.166604} in order to explain a remarkable value of $g_{h2}$ observed experimentally we 
surmised that the $\pm 3/2$ heavy hole mixing could occur due to a cubic-in-${\bm k}$ contribution ${\cal H}_{v3}$ to the hole Hamiltonian. However, the subsequent estimation has shown that the corresponding constant ${\cal A}^{(3)}$ is too small to be responsible for the effect. Here we propose a mechanism of the heavy hole mixing which is based on the standard Luttinger Hamiltonian quadratic in ${\bm k}$ and an anisotropic shape of the GaAs/AlGaAs(111) QDs including polar orientation along the growth direction $z$ and trigonal symmetry in the plane perpendicular to $z$.
\subsubsection{Quantization of the hole motion in the pyramidal quantum dot}
The hole states in a QD are described in the
  framework of the Luttinger Hamiltonian $\mathcal{H}_{\Gamma_8} +
  \mathcal{V}({\bm r})$, where $\mathcal V(\bm r)$ is the operator of
  potential energy of the hole and $\mathcal{H}_{\Gamma_8}$ is written in the standard matrix form\cite{ivchenko05a} 
\begin{equation}
\label{eq:Lut}
\mathcal H_{\Gamma_8} =
\begin{pmatrix}
F & H & I & 0 \\  
H^* & G & 0 & I \\
I^* & 0 & G & -H \\  
0 & I^* & -H^* & F\\
\end{pmatrix}.
\end{equation}
In the spherical approximation used hereinafter, the functions $F$, $G$, $H$, and $I$   read
\[
F = (A-B) k_z^2 + \left( A + \frac{B}{2} \right) \left( k_x^2 + k_y^2 \right),
\]
\[
G = (A+B) k_z^2 + \left( A - \frac{B}{2} \right) \left( k_x^2 + k_y^2 \right),
\]
\[
H = -\sqrt{3}B{k}_z\left( {k}_x - {\rm i}{k}_y \right),
\quad I = -\frac{\sqrt{3}}{2} B \left( {k}_x - {\rm i} {k}_y \right)^2 .
\]
Here $k_j$ ($j=x,y,z$) are the Cartesian components of wavevector (we remind that the axes frame used is  $x \parallel [11\bar{2}]$, $y \parallel
[\bar{1}10]$ and $z \parallel [111]$). We will further use hole representation so that constants $A$ and $B$ are related to the Luttinger parameters $\gamma_1$ and $\gamma_2 = \gamma_3 \equiv \bar{\gamma}$ as $A = \hbar^2 \gamma_1/ 2m_0$, $B = \hbar^2 \gamma_2/ m_0$,  $m_0$ being the free electron mass. 
In the effective Hamiltonian~\eqref{eq:Lut} the wavevector $\bm k$ is replaced by the operator $-\mathrm i \nabla$.   

We take the potential in the following matrix form
\[
\mathcal V(\bm r) = \begin{pmatrix}
V_{hh}(\bm r) & 0 & 0 & 0 \\
0 & V_{lh}(\bm r) & 0 & 0 \\
0 & 0 & V_{lh}(\bm r)  & 0 \\
 0 & 0 & 0 & V_{hh}(\bm r) 
\end{pmatrix},
\]
allowing for different potential energies of the heavy {(}$V_{hh}${)} and {light} {(}$V_{lh}${)} holes.
In self-organized QDs the size quantization along the growth axis $z$ is stronger than that in the dot plane allowing
one to separate the hole motion along $z$ and in the $(x,y)$ plane. As a reasonable simplification, the potentials acting on each kind of holes, $n=hh$ or $n=lh$, are modelled by separable functions,
\begin{equation}
\label{eq:potential}
V_n(\bm r)  = V^{n}_z(z) + V^{n}_{\parallel}(\rho, \varphi),
\end{equation}
where for convenience we use the cylindrical coordinate system with
the principal axis $z$ and in-plane polar coordinates $\rho=\sqrt{x^2+y^2}$ and
$\varphi$.  
The eigenstates and eigenenergies are found from the time-independent Schr\"{o}dinger equation
\[
[\mathcal H_{\Gamma_8} + \mathcal V(\bm r)] \hat \Psi(\bm r) = E\hat \Psi(\bm r),
\]
where $\hat \Psi(\bm r)$ is a column formed by four envelopes $\Psi_m(\bm r)$ with $m=-3/2,-1/2,1/2,3/2$.

The heavy-light hole mixing is described by the off-diagonal components of the Luttinger Hamiltonian~\eqref{eq:Lut}. 
While calculating the hole $g$-factor, we take into account this mixing in the first-order approximation. In the zeroth-order approximation, however, the off-diagonal components of $\mathcal H_{\Gamma_8}$ can be  neglected at all, thereby, heavy- and light-hole states are completely separated in this approximation, so that the four-component hole wavefunction $\hat \Psi$ has only one non-zero component $\Psi_m^n$ with $m=-3/2$ either $3/2$ for $n=hh$ and $m=-1/2$ either $1/2$ for $n=lh$.\cite{semina_suris_2011}
  Due to the assumed separable potential~\eqref{eq:potential} the envelopes $\Psi_m^n(\bm r)$ can be sought in separable form\cite{bastard83,wojs96b}
\begin{equation} \label{Psinmlp}
\Psi_{m;lp}^n(\bm r) = F_l^n(z) \psi^n_p(\rho,\varphi)\:.
\end{equation}
Here $F^n_l(z)$ describes the hole size-quantization along the growth axis, with $l=1,2\ldots$ is an integer enumerating the
size-quantized states along this axis, and $\psi^n_p(\rho,\varphi)$ is the in-plane envelope with $p=1,2\ldots$ labeling the in-plane confined states. 

In the further specification of the potential~\eqref{eq:potential} we bear in mind that the QDs under consideration 
have the approximate shape of triangular pyramids asymmetric with respect
to the $z \to - z$ reflection and possess the three-fold rotation symmetry in the $(x,y)$ plane, see Fig.~\ref{fig:fig1} and Ref.~\onlinecite{Mano2010}. The asymmetry along the growth axis is modelled
by a triangular potential well,
\begin{equation}
  \label{eq:Vz}
  V_z^n(z) = \begin{cases} -e \mathcal{F} z,& z>0 \\ +\infty,& z<0 \end{cases}\:,
\end{equation}
where $z=0$ corresponds to the QD base, $\mathcal{F}$ is the
effective electric field breaking the $z\to -z$ mirror symmetry and confining a
hole along the $z$-axis, and $e=-|e|$ is the electron charge. In this case the $z$-dependent
envelopes are expressed via the Airy functions as follows
\begin{equation}
\label{Fz}
F_{l}^n (z) = C_{l}^n {\rm Ai}(Z_l^n)\:,
\end{equation}
where $C^n_l$ is the normalization factor,
\[
Z^n_l = \frac{z}{L_n} - \mu_l \quad,\quad L_n = \left( \frac{\hbar^2}{2m_{n,z}e\mathcal{F}} \right)^{1/3}\:, 
\]
$\mu_l$ is the $l$-th root of the equation ${\rm Ai}(-Z) = 0$, and the effective mass $m_{n,z}$ equals to
$m_0/(\gamma_{1} \pm 2 \bar{\gamma})$ with the $\pm$ sign corresponding to the light and heavy holes. 

The in-plane potential acting on holes is taken to be an analytical function of ${\bm \rho} $ and comprise the main parabolic term~\cite{que92,hawrylak99,semina_2006} and a cubic correction describing the trigonal warping, namely, 
\begin{equation}
\label{eq:Vinplane}
V^n_\parallel(\rho,\varphi) = \frac{\hbar^2 \rho^2}{2m_{n,\parallel}
  a_n^4}  \left( 1 + \beta \frac{\rho}{a_n} \cos{3\varphi} \right)\:. 
\end{equation} 
Here $m_{n,\parallel} = m_0/(\gamma_1 \pm \bar{\gamma})$ are the effective masses for
    heavy (top sign) and light (bottom sign) holes in the QD
    plane,
    $a_n$ are the effective localization radii and the
    dimensionless parameter $\beta$ characterizes the quantum dot
    triangularity. 
    
\begin{figure}[hptb]
\includegraphics[width=0.2\textwidth]{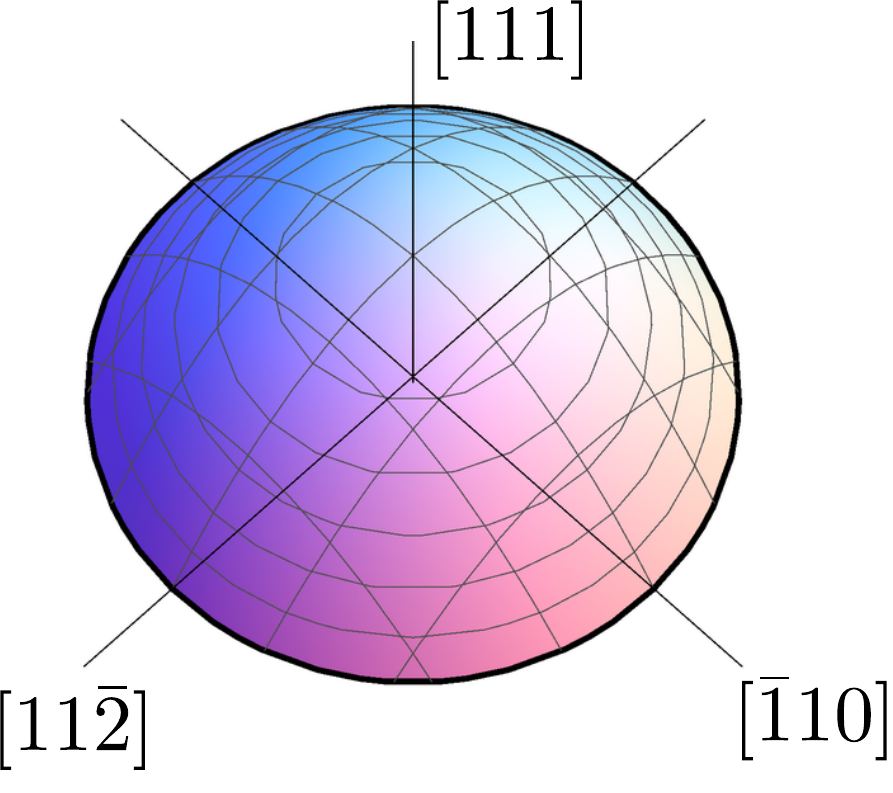} (a)
\includegraphics[width=0.2\textwidth]{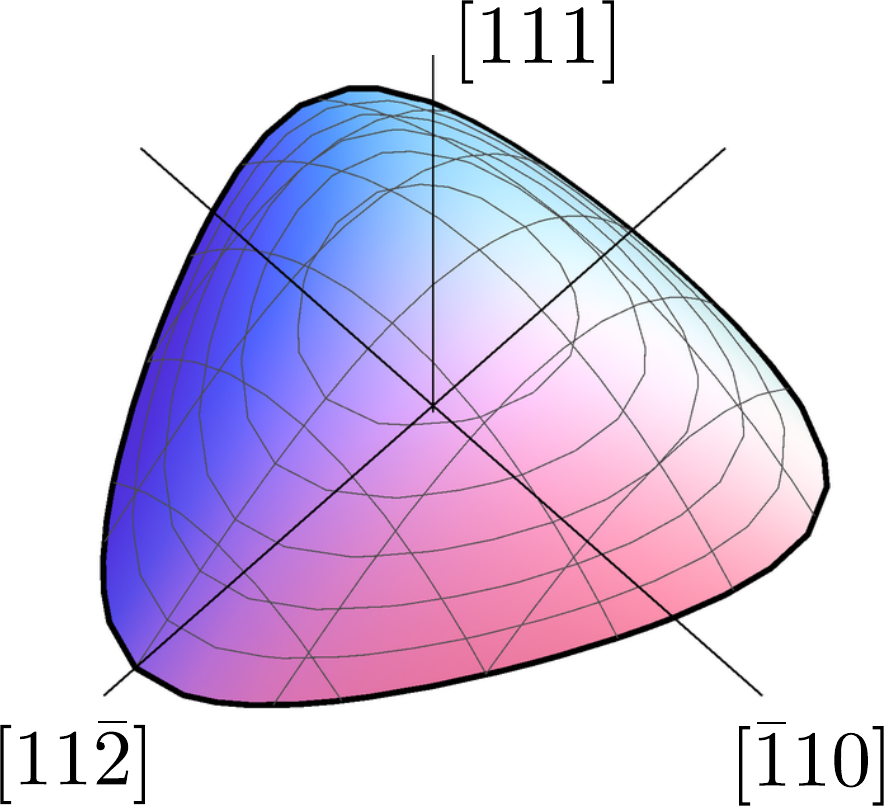} (b)
\caption{Equipotential surface of the dot confining potential
for the two values of triangularity parameter $\beta  = 0$ (a) and $\beta = -0.35$ (b)}
\label{fig:pot3D}
\end{figure}

    Figure~\ref{fig:pot3D} represents equipotential surfaces of the dot
confining potential defined by Eqs.~\eqref{eq:potential}, (\ref{eq:Vz}), and (\ref{eq:Vinplane}) for $\beta=0$ (a) and $\beta=-0.35$ (b) and illustrates the transfer from the axial to trigonal symmetry.


For the GaAs parameters $\gamma_1 = 6.98$ and $\bar{\gamma} = 2.58$, 
one has $m_{hh,z} = 0.55 m_0$, $m_{lh,z} =
0.08 m_0$ $m_{hh,\parallel} = 0.1m_0$, and $m_{lh,\parallel} =0.22m_0$. The size-quantization energies for the motion along
  $z$-axis are given by
\begin{equation}
\label{Ez}
E_{z,l}^{n} =  \frac{\hbar^2 \mu_l}{2m_{n,z} L_n^2}\:,\hspace{3 mm} l= 1,2\ldots
\end{equation}
It is convenient to introduce the dot effective size in the 
$z$-direction as $L = \sqrt{L_{hh} L_{lh}}$. For $L=30$~\AA~the values of energies of the ground heavy and light-hole states are, respectively, $E_{z,1}^{hh} \approx 33$ meV and $E_{z,1}^{lh} \approx 63$ meV. The energies of size quantization in the quantum dot plane are labeled as $E_{\parallel,p}^{hh}$ and $E_{\parallel,p}^{lh}$ for heavy and light holes, respectively. If $\beta=0$ the energies of in-plane size-quantization form an equidistant set with the spacing
  \begin{equation}
    \label{eq:equidistant}
    \hbar\omega_n = {\frac{\hbar^2}{m_{n,\parallel}a_n^2}}\:,
  \end{equation}
and the eigenfunctions are those of a two-dimensional
  axially-symmetric harmonic oscillator. 
In what follows we assume
  that the confining in-plane potential in the case of $\beta = 0$ is
  identical for the heavy and light holes. This implies that
  $m_{hh,\parallel}a_{hh}^4 = m_{lh,\parallel}a_{lh}^4$ and
  the ratio of two localization radii for GaAs is $a_{hh}/a_{lh}
  \approx 1.21$. Taking  $75$~\AA~as a reasonable estimation for $a_{hh}$ we obtain  $\hbar \omega_{hh} \approx 6.5$~meV and $\hbar
  \omega_{lh} \approx 4.4$~meV. These values are small as compared to the energy difference $E_{z,1}^{lh}-E_{z,1}^{hh} \approx 30$ meV of hole confinement in the $z$-direction, confirming the consistency of the model. 
  
\subsubsection{Longitudinal $g$-factor $g_{h1}$}
The Zeeman interaction of a bulk $\Gamma_8$ hole with the
magnetic field $\bm{B}$ is described by the $4\times 4$ matrix
operator~\cite{birpikus}, resulting for ${\bm B} \parallel [111]$ in the following expressions
for the effective $g$-factors:~\cite{PhysRevLett.107.166604} 
\begin{subequations}
\label{eq:gbulk}
\begin{equation}
\label{bulk:gh1}
g_{h1} = -6\varkappa - \frac{23}{2} q, 
\end{equation}
\begin{equation}
\label{bulk:gh2}
g_{h2} = 2\sqrt{2} q,
\end{equation}
\end{subequations}
where  $\varkappa$ and $q$ are the dimensionless
band structure parameters, describing isotropic and cubic
contributions to Zeeman effect.\cite{luttinger_1956, PhysRev.133.A542} In
GaAs crystals $\varkappa = 1.2$, $q=0.02$.~\cite{Mar99} It follows from
Eq.~(\ref{bulk:gh2}) that the off-diagonal $g$-factor $g_{h2}$ is
already present in this model, however, its value $g_{h2} \approx
0.056$ is too small to describe the experiment. Moreover, similarly to bulk
semiconductors and quantum 
wells,\cite{luttinger_1956, ram_mohan_1988, PhysRevB.50.8889, kiselevmoiseev96_eng} the magnetic field
induced mixing of heavy and light hole states and quantum confinement strongly renormalizes
heavy-hole $g$-factor in quantum dots from its value in the
bulk.\cite{PhysRevB.67.205301, Nakaoka2004, Nakaoka2005, Pryor2006, PhysRevB.79.045307,
    Sheng2009, Prado2004} 
Indeed, in the presence of the
magnetic field the wavevector $\bm k$ in Eq.~(\ref{eq:Lut})
is replaced by $\bm k + e\bm A(\bm r)/c\hbar$, where $\bm A(\bm
r) = (1/2) {\bm B} \times {\bm r}$ is the vector potential of the magnetic field.\cite{hole:rep} The vector potential, through the off-diagonal components of the matrix (\ref{eq:Lut}), mixes the ground heavy-hole state with light-hole states. Calculation presented in  Appendix~\ref{app:gh1} shows that the longitudinal $g$-factor  $g_{h1}$ in the quantum dot is given by
\begin{equation} \label{gh1delta}
g_{h1} = -6\varkappa - (23/2) q + \Delta g_{h1}\:,
\end{equation}
where the third term obtained in the second order in $L/a$ has the form 
\begin{equation}
\label{eq:gh1}
\Delta g_{h1} = 24\bar{\gamma}^2 \frac{\sqrt[3]{m_{hh,z} m^2_{lh,z}}}{m_0}
 \sum_l \frac{{\eta_l^2} + {\xi_l^2} (L^2/2a^2)}{\mu_l - \mu_1
   \sqrt[3]{m_{lh,z}/m_{hh,z}}}\:.
\end{equation}
Here $a \equiv a_{hh}$, other notations are
\begin{eqnarray}
{\eta_l}&=& L\int   F_{1}^{hh}(z)\frac{\partial}{\partial z} F_{l}^{lh}(z) dz\:, \nonumber
\\ {\xi_l^2} & =& {\zeta_l^2 - \frac{4 \eta_l^2}
{  \mu_l \sqrt{m_{lh,\parallel} m_{hh,\parallel}}/m_{lh,z} - \mu_1
    m_{hh,\parallel}/m_{hh,z} } } \:, \nonumber \\
\zeta_l &=& \int F_{1}^{hh}(z) F_{l}^{lh}(z) dz\:.
\end{eqnarray}  

Performing summation in Eq.~(\ref{eq:gh1}) over $l$ for the set of GaAs parameters $\gamma_1$, $\bar{\gamma}$, $m_{hh,\parallel}$ and $m_{lh,\parallel}$ we obtain for $g_{h1}$
\begin{equation}
\label{gh1_est}
g_{h1} \approx -6\varkappa - (23/2) q + {4.5} + {6.7}\frac{L^2}{a^2}\:,
\end{equation}
where the third and fourth coefficients are calculated numerically.
For the $g_{h1}$ calculations the in-plane anisotropy of the
dot was neglected, i.e., in the framework of our model we set
$\beta =0$ and chose the angular envelopes $\psi_{p}^n$ [see
Eq.~(\ref{Psinmlp})] as the eigenfunctions of
isotropic two-dimensional harmonic oscillator.
Note, that trigonal distortion of the in-plane potential described by
a small parameter $\beta$ leads to a correction to $g_{h1}$
quadratic in $\beta$. Indeed, as it
  follows from the symmetry
  reasons the change of the $\beta$ sign is equivalent to the
  rotation of the triangular QD around $z$ axis by $\pi$, in
    which case $g_{h1}$ remains unchanged. Estimations show
that this quadratic correction is negligible.   

\subsubsection{Off-diagonal longitudinal g-factor $g_{h2}$}
The pyramidal shape of the quantum dot with the three-fold rotation axis $z$ results in the magnetic-field induced mixing of heavy holes with opposite angular momentum projections $\pm 3/2$. 
The trigonal distortion of the in-plane parabolic confinement  described by the term $\propto \beta \rho^3 \cos 3\varphi$ in Eq.~(\ref{eq:Vinplane}) leads to the lowering of symmetry of
the ground heavy-hole state envelope function, which can be represented in the lowest order in $|\beta| \ll 1$ as
\begin{equation}
\label{eq:psi00trig}
\tilde{\psi}_{1}^{hh} \propto \exp\left( -\frac{\rho^2}{2a^2}
\right)\left( 1 + \beta \frac{\rho^3}{a^3}C(\rho)\cos
  3\varphi \right),
\end{equation}
where function $C(\rho)$ is finite at $\rho=0$ and only
terms up to linear order in $\beta$ are retained. The variational procedure with the space-independent $C(\rho) \equiv C_0$ gives for $C_0$ a value of $-1/6$.

By means of the second-order
perturbation theory we obtain for the magneto-induced mixing of $3/2$ and
$-3/2$ heavy-hole states [cf. Eq.~(\ref{2x2})]:
\begin{eqnarray}
\label{eq:mixing}
\mathcal{H}_{B, 3/2, -3/2} \equiv \frac12 \mu_B g_{h2} B_z = \hspace{2 cm} \\
\sum_{lp\pm} \frac{\left\langle \Psi^{hh}_{3/2;11}| \hat{\mathcal{H}}_
{\Gamma_8} |\Psi^{lh}_{\pm 1/2; lp} \right\rangle \left\langle \Psi_{\pm 1/2;lp}^{lh}|
     \hat{\mathcal{H}}_{\Gamma_8} |\Psi_{-3/2;11}^{hh} \right\rangle}
 {E_{z,1}^{hh} + E_{\parallel,1}^{hh}  - E_{z,l}^{lh} -
   E_{\parallel,p}^{lh}}. \nonumber
\end{eqnarray} 
We note that while replacing in $\mathcal H_{\Gamma_8}$ the hole wavevector $\bm k$  by $\bm k + e\bm A/\hbar c$ only linear in $\bm A$ terms should be retained.

The details of evaluation of the off-diagonal parameter $g_{h2}$ are presented in
Appendix~\ref{app:gh2}. Similarly to the calculation of $g_{h1}$, in the limit $L/a \ll 1$ 
one can neglect the dependence of hole energy on the
in-plane quantum numbers $p$ with the result
\begin{multline}
\label{gh2_simplified}
g_{h2} = \frac{18 \hbar^2 \bar{\gamma}^2}{m_0 L} \sum_{l}  \frac{\eta_l
  \zeta_l}{E^{lh}_{z,l} - E^{hh}_{z,1}}  \\ 
\times  \left\langle \tilde{\psi}_1^{hh} \left| (x - {\rm i}y) \left( \frac{\partial}{\partial x} - {\rm i} \frac{\partial}{\partial y} \right)^2 \right| \tilde{\psi}_{1}^{hh}\right\rangle \:.
\end{multline}
This equation clearly shows that the value of $g_{h2}$ is nonzero provided only that 
(i) the dot base has a triangular shape, i.e., $\beta \neq 0$, otherwise integration over the in-plane
coordinates vanishes, and (ii) the dot is asymmetric in the $z$ direction, otherwise 
the product $\eta_l \zeta_l$ vanishes due to parity. 

Using the variational function (\ref{eq:psi00trig}) with $C(\rho) = - 1/6$ we obtain
\begin{multline}
\label{gh2_2}
g_{h2} = 36 \beta \bar{\gamma}^2 \frac{\sqrt[3]{m_{hh,z} m^2_{lh,z}}}{m_0}
\frac{L}{a}  \\ 
\times \sum_l \frac{\eta_l \zeta_l}{\mu_l - \mu_{1}
  \sqrt[3]{m_{lh,z}/m_{hh,z}}}\:. 
\end{multline}
The numerical summation in Eq.~(\ref{gh2_2}) yields
\begin{equation} \label{gh2betaLa}
g_{h2} \approx 10.2 \beta \frac{L}{a}\:,
\end{equation} 
where the small contribution $2\sqrt{2}q$ is neglected.
For reasonable values $\beta = -0.2$, $L/a = 0.3$ we obtain
$|g_{h2}| \approx 0.62$, in good agreement with experiment.

\section{Discussion}
\begin{figure}[hptb]
\includegraphics[width=0.47\textwidth]{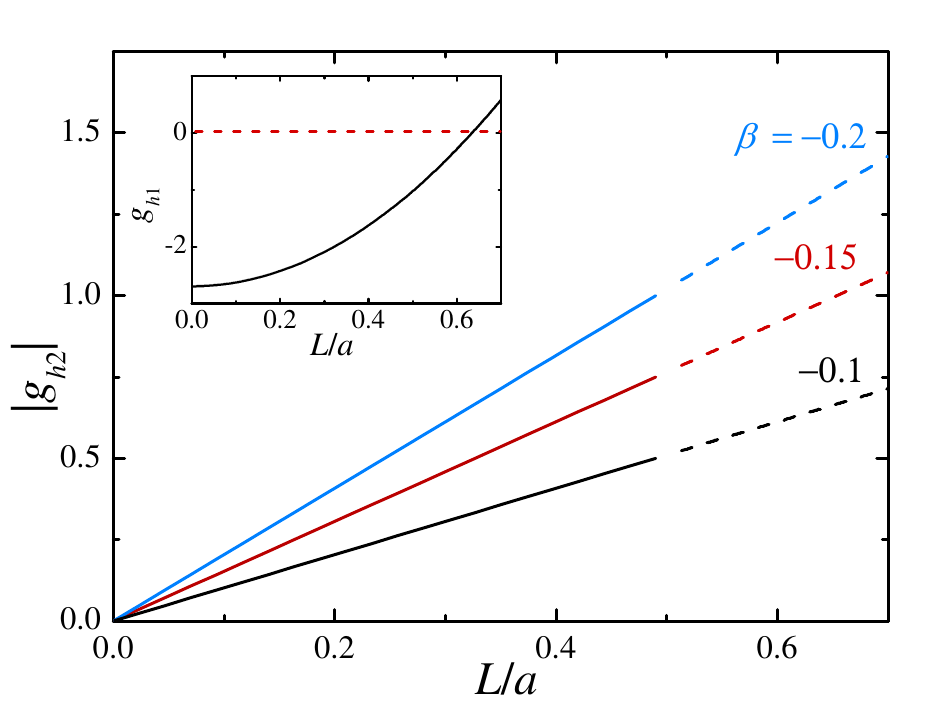} 
\caption{The off-diagonal $g$-factor $g_{h2}$ as a function of $L/a$ depicted for three values of the trigonal parameter $\beta$: $-0.1$, $-0.15$ and $-0.2$. The inset shows the behavior of $g_{h1}$.}
\label{fig:size}
\end{figure}

Figure~\ref{fig:size} displays the $g$-factors $g_{h1}$ and $g_{h2}$ calculated as a function of the ratio $L/a$ according to Eqs.~(\ref{gh1_est}) and (\ref{gh2betaLa}). The figure shows that the values of $|g_{h2}|$
vary in the range of $0.3 \div 0.7$ for a reasonable range of the parameter $\beta$ and the ratio $L/a$ in good agreement with the experimental observations. The diagonal $g$-factor, $g_{h1}$, changes its sign at some critical value of
$L/a$ which depends on the material parameters and quantum dot shape. We note that for the particular form of the Zeeman Hamiltonian, Eq.~\eqref{2x2}, the sign of $g_{h1}$ cannot directly be extracted from the PL data in the magnetic field. Relying on our estimations we believe, however, that the value of $g_{h1}$ for studied dots
with small values of $L/a$ is negative.

Accurate numerical estimations of effective $g$-factors require the knowledge of spacing between size-quantized levels of heavy and light holes in the $z$-direction. The value of this spacing can be estimated using the anisotropy of the electron $g$-factor, i.e., the difference between its longitudinal ($g_{e,\parallel}$) and transverse ($g_{e,\perp}$) components. It is well known that the $g$-factor of an electron confined in a quantum dot is strongly anisotropic,\cite{PhysRevB.75.245302, PhysRevB.58.16353, ivchenko05a} and the value of that anisotropy is directly related to the energy spacing between heavy- and light-hole states.
Taking for the experimental values of $g_{e,\parallel}$ and $g_{e,\perp}$ their relative difference $(g_{e,\perp} - g_{e,\parallel})/g_{e,\parallel} \approx 40\%$ and using formulae of Ref.~\onlinecite{ivchenko05a}, the energy spacing $E^{lh}_{z, 1} - E^{hh}_{z, 1}$ can be estimated as $20 \div 30$~meV, which is in a good agreement with the values obtained in our simple model of the confinement in the $z$-direction with $L$ ranging from 25 to 35~\AA.

\begin{figure}[hptb]
\includegraphics[width=0.47\textwidth]{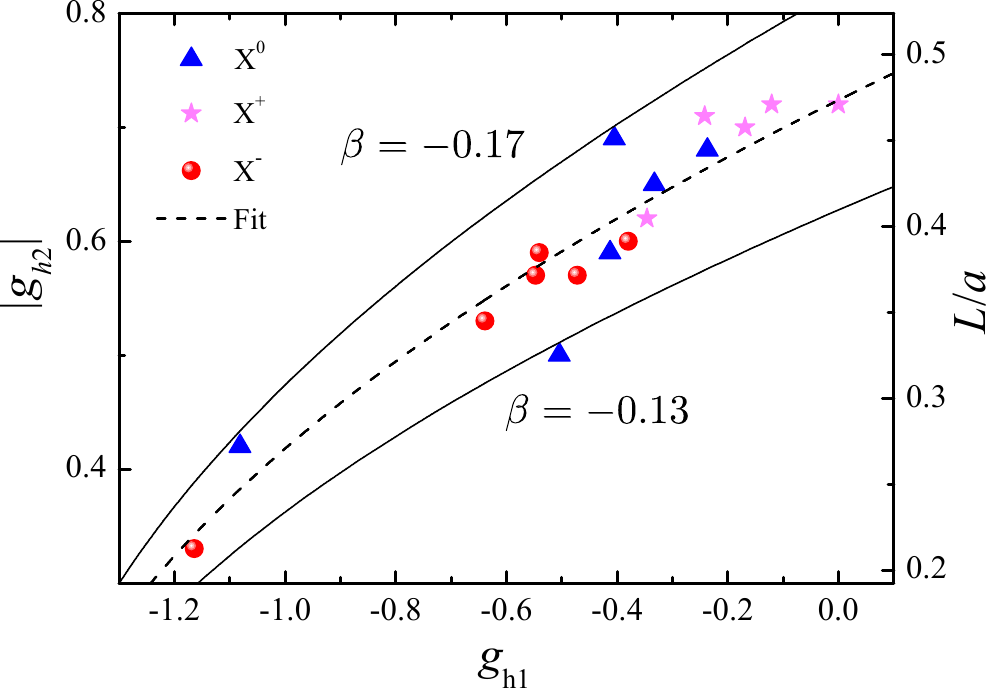} 
\caption{Correlation between the values of $g_{h1}$ and $|g_{h2}|$. Symbols represent experimental data for different GaAs dots measured on the neutral as well as on charged complexes, dashed curve stands for theoretical fit with parameters $\beta = -0.15$, $g_{h1}^0 =-1.5$, $c_1 =6.7$, $c_2 = 10.2$. Thin lines are calculated for the same set of parameters, except for $\beta=-0.13$ and $-0.17$. The corresponding values of $L/a$ (for $\beta =-0.15$) are presented at the right-hand axis.}
\label{fig:correl}
\end{figure}

It should be noted that Eqs.~(\ref{gh1_est}) and (\ref{gh2betaLa}) are derived assuming $L/a \ll 1$. We presume that they can reasonably be extrapolated up to QDs with $L/a$ of the order of unity and use the following working equations
\begin{eqnarray} \label{gh1gh2}
g_{h1}(L/a) &=& g_{h1}^0 + c_1 \frac{L^2}{a^2}\:, \nonumber \\
g_{h2}(L/a) &=& c_2 \beta \frac{L}{a}\:,
\end{eqnarray}
where $g_{h1}^0 = g_{h1}(0)$ and the coefficients $c_1, c_2$ are considered as fitting parameters. The absence of zero-order term in the dependence of $g_{h2}$ on $L/a$ is expected since, in the band model under consideration, the heavy-hole mixing in quantum wells is negligible. In contrast, the diagonal component $g_{h1}$ is strongly renormalized even in quantum wells. The quadratic term in (\ref{gh1_est}) appears due to the small term $E^{hh}_{\parallel, 1} -
E^{lh}_{\parallel, p}$ in the denominators of Eq.~(\ref{dEB}). 
Eliminating the variable $L/a$ we directly relate the diagonal and off-diagonal components of the $g$-factor as
\begin{equation}
\label{eq:correl}
g_{h2} = c_2 \beta \sqrt{\frac{g_{h1} - g_{h1}^0}{c_1}}\:.
\end{equation}
The product $c_2 \beta/\sqrt{c_1}$ can be considered an indivisible parameter. We, however, take $c_1=6.7$, $c_2=10.2$ as follows from Eqs.~\eqref{gh1_est}, \eqref{gh2betaLa} and fit the value of $\beta$ to the experimental data. Equation (\ref{eq:correl}) predicts a particular correlation between the diagonal and off-diagonal hole $g$-factors.
Figure~\ref{fig:correl} represents experimental results as well as theoretical fit with parameters introduced in the caption. As mentioned above, the experimental values of $g_{h1}$ are taken with negative signs. One can see that the theory well reproduces
the experimental relation between $g_{h2}$ and $g_{h1}$.

The systematic differences between the off-diagonal $g$-factors $g_{h2}$ for $X^+$ and $X^-$ trions seen in Table~\ref{tab:table1} indicate a non-negligible role of the Coulomb interaction. Indeed, the value of $g_{h2}$ found from the spectrum of $X^+$ trion radiative recombination corresponds to the off-diagonal $g$-factor of a resident hole, while the value $g_{h2}$ extracted from the $X^-$ spectrum gives the $g$-factor of the hole in a trion. However, the detailed investigation of Coulomb-driven effects is beyond the scope of the present paper.


As far as we use Hamiltonian Eq.~(\ref{eq:Lut}) in the spherical approximation, the [111] growth direction is nominally equivalent to [001] direction, except for natural trigonal shape of the confining potential of QDs grown in the former case \cite{Mano2010,Treu:2012a}. A nonzero value of $g_{h2}$ is also obtained if we neglect the QD triangularity but make allowance for a cubic anisotropy of the Luttinger Hamiltonian, described by the nonvanishing difference $\delta \gamma = \gamma_2 - \gamma_3$ which is a small parameter for majority of semiconductor materials. Transformation of this Hamiltonian from the crystallographic axes [100], [010], [001] to the frame $x$,$y$,$z$
results in its matrix form (\ref{eq:Lut}) where $H$ and $I$ acquire corrections 
\begin{eqnarray}
\label{eq:cubic}
\delta H &=& -\frac{\hbar^2 \delta \gamma}{\sqrt{6}m_0} \left( k_+^2 - 2\sqrt{2}k_{z} k_- \right)\:, \nonumber \\
\delta I &=& \frac{\hbar^2 \delta \gamma}{2\sqrt{3}m_0} \left( k_-^2 - 2\sqrt{2}k_{z} k_+ \right)\:,
\end{eqnarray}
where $k_{\pm} = k_{x} \pm {\rm i} k_{y}$.
These corrections lead to nonzero values of $g_{h2}$ even for the disk- and lens-shaped dots as well as for the [111] grown quantum wells. However, the corresponding contribution $\delta g_{h2} \propto \delta \gamma$ is indeed small and, in the case of trigonal dots, constitutes a fraction $\leq$ 10\% of that obtained from Eq.~(\ref{gh2_2}).
As mentioned above, an additional mechanism of the magneto-induced mixing of the $\pm 3/2$ hole states suggested in Ref.~\onlinecite{PhysRevLett.107.166604} and based on a cubic-${\bm k}$ contribution to the hole Hamiltonian also gives too small value of $|g_{h2}|$. 

Therefore, we can conclude that a sizable off-diagonal heavy-hole $g$ factor, $g_{h2}$, indeed results from the quantum dot shape in the form of triangular pyramid. Note that this is independently confirmed by an absence of the heavy-hole mixing in [111] grown disk-shaped dots studied experimentally in Ref.~\onlinecite{PhysRevB.84.195305}.
\section{Conclusions}
In this paper we have shown that GaAs droplet dots grown along $[111]$ axis demonstrate striking magneto-photoluminescence spectra. In longitudinal field $\bm B \parallel [111]$, each electron-hole complex, a neutral exciton X$^0$ and positively X$^+$ or negatively X$^-$ charged exciton, shows four emission lines with definite circular polarizations, instead of two lines for commonly studied (001) dots. Moreover, in transverse field $\bm B \perp [111]$ only two linearly polarized lines are observed. These effects are described by the magneto-induced mixing of the heavy holes for $\bm B \parallel [111]$ and the absence of such mixing for the transverse field.

A microscopic theory of the mixing has been developed in the framework of Luttinger Hamiltonian approach adopted to describe the hole quantization in the dots with a shape of triangular pyramid. Our theory is in good quantitative agreement with experimental data. We have established that it is the shape of the dot which results
in off-diagonal components of the $g$-factor tensor and gives rise to the novel magneto-optical phenomena in droplet (111) quantum dots.

\acknowledgements

Financial support from RFBR, RF President Grant
NSh-5442.2012.2, EU projects SPANGL4Q, ITN SpinOptronics, POLAPHEN, ANR QUAMOS and LIA ILNACS is gratefully acknowledged.

\appendix
\section{Calculation of $g_{h1}$}\label{app:gh1}
The additional term $\Delta g_{h1}$ in Eq.~(\ref{gh1delta}) for the diagonal $g$-factor $g_{h1}$ is found in the second order of perturbation theory from the following relation
\begin{eqnarray} \label{dEB}
\frac12 \Delta g_{h1} \mu_B B_z = \hspace{3 cm}\\
\sum_{lp \pm} \frac{\left\langle \Psi_{3/2;11}^{hh}| \hat{\mathcal{H}}_{\Gamma_8} |\Psi_{\pm 1/2;lp}^{lh} \right\rangle \left\langle \Psi_{\pm 1/2;lp}^{lh}| \hat{\mathcal{H}}_{\Gamma_8} |\Psi_{3/2;11}^{hh} \right\rangle}{E_{z,1}^{hh} + E_{\parallel,1}^{hh}  - E_{z,l}^{lh} - E_{\parallel,p}^{lh}}\:, \nonumber
\end{eqnarray}
where the summation includes all  the light hole states described by the two indices $l$ and $p$. Using the hole envelopes one can rewrite~\eqref{dEB} in the form 
\begin{widetext}
\begin{eqnarray}
\label{dEB2}
\Delta g_{h1} &=&  (2/\mu_B B_z)\sum_{lp}  (E_{z,1}^{hh} + E_{\parallel,1}^{hh}  - E_{z,l}^{lh} -
   E_{\parallel,p}^{lh})^{-1}
\\ && \times \left( \left\langle F_{1}^{hh} \psi_{1}^{hh}| \tilde{H}
      |F_{l}^{lh} \psi_{p}^{lh}\right\rangle \left\langle F_{l}^{lh}
      \psi_{p}^{lh}| \tilde{H}^* | F_{1}^{hh} \psi_{1}^{hh}\right\rangle +
    \left\langle F_{1}^{hh} \psi_{1}^{hh}| \tilde{I} |F_{l}^{lh}
      \psi_{p}^{lh} \right\rangle \left\langle F_{l}^{lh} \psi_{p}^{lh}|
      \tilde{I}^* |F_{1}^{hh} \psi_{1}^{hh} \right\rangle \right)\:. \nonumber
\end{eqnarray} 
\end{widetext}
Here $\tilde{H}$ and $\tilde{I}$ are the components of
the Hamiltonian~(\ref{eq:Lut}) modified in the presence of
magnetic field: 
\[
\tilde{H} = -\sqrt{3}B k_z\left[k_x - {\rm i} k_y + \frac{e}{c\hbar}(A_x - {\rm i}A_y) \right]\:,
\]
\[
\tilde{I} = -\frac{\sqrt{3}}{2}B\left[(k_x - {\rm i}k_y)^2  + \frac{2e}{c\hbar}\left\{(k_x - {\rm i}k_y)(A_x - {\rm i}A_y)\right\} \right]\:,
\] 
$k_{\alpha}$ should be replaced by the differential operators $- {\rm i} \partial/ \partial x_{\alpha}$, the curly brackets denote a symmetrized product of the corresponding operators.
In the calculation of the sum (\ref{dEB2}) one has to retain only linear
contributions in ${\bm A}$. In the limit $L/a \ll 1$ the energy denominator in Eq.~(\ref{dEB2}) can be expanded in powers of this ratio. Performing summation over the index $p$ we directly arrive at Eq.~(\ref{eq:gh1}) from the main text.

\section{Calculation of $g_{h2}$}\label{app:gh2}
Starting from Eq.~(\ref{eq:mixing}) and using the components of the Luttinger Hamiltonian $\tilde{H}, \tilde{I}$ in the magnetic field we arrive at 
\begin{multline} \label{gh2}
g_{h2} = \frac{12 \hbar^2 \bar{\gamma}^2}{m_0 L} \sum_{lp}  \frac{\eta_l \zeta_l}{E_{z,1}^{hh} + E_{\parallel,1}^{hh}  - E_{z,l}^{lh} - E_{\parallel,p}^{lh}} 
\\ 
 \times 
\left(\frac{1}{2} \left\langle \tilde{\psi}_{1}^{hh} \left| \nabla_{-}^2 \right|\psi_{p}^{lh} \right\rangle \left\langle \psi_{p}^{lh} \left| r_{-} \right| \tilde{\psi}_{1}^{hh}\right\rangle 
\right. \\
\left.
+ \left\langle \tilde{\psi}_{1}^{hh} \left| \nabla_{-} \right| \psi_{p}^{lh} \right\rangle \left\langle \psi_{p}^{lh} \left| \left\{ \nabla_{-} r_{-} \right\} \right|  \tilde{\psi}_{1}^{hh}  \right\rangle \right)\:,
\end{multline}
where  $r_{-} = x - {\rm i} y \propto {\rm e}^{-{\rm i} \varphi}$, $\nabla_{-} = \partial / \partial x- {\rm i}\partial/\partial y \propto {\rm e}^{-{\rm i} \varphi}$, $\nabla_{-}^2 \propto {\rm e}^{-2 {\rm i} \varphi}$. Note that at $\beta = 0$ the mixing described by Eq.~(\ref{gh2}) vanishes because in this case the matrix elements of operators proportional to ${\rm e}^{-{\rm i} \varphi}$ and ${\rm e}^{-2{\rm i} \varphi}$ cannot be nonzero simultaneously.

Retaining the first-order term in the expansion of $g_{h2}$ in powers of $L/a \ll 1$ we can neglect the in-plane energies $E_{\parallel, p}^{h}$ in the denominators of Eq.~(\ref{gh2}). Then the summation in Eq.~(\ref{gh2}) over the index $p$ can be performed analytically, and we directly arrive at Eq.~(\ref{gh2_simplified}).


\begin{thebibliography}{80}
\bibitem{aniso111} {The anisotropic short-range exchange may result in the dark-bright mixing for (111) quantum dots even in the absence of magnetic field.\cite{PhysRevLett.107.166604} Estimates show, however, that this effect is quite minor, correspondingly, no experimental evidence of such a mixing is observed.}
\bibitem{hole:rep} Here we use the hole representation.
\end{thebibliography}
\end{document}